\documentclass{jltp}
\usepackage{graphicx}
\usepackage{bm} 
\usepackage{epsf}
\title{Simulations of Dynamical Ordering in Pinned Vortex Systems}

\author{M. F. Laguna, P. S. Cornaglia, and C. A. Balseiro}

\address{Centro At\'{o}mico Bariloche, 8400 San Carlos de Bariloche, R\'{\i}o Negro, Argentina}

\runninghead{M. F. Laguna {\it et al.}}{Dynamical ordering in vortices}

\begin{document}

\maketitle

\begin{abstract}
We model a vortex system in a sample with bulk pinning and superficial pinning generated by a magnetic decoration. We perform a sequence of finite temperature numerical experiments in which external forces are applied to obtain a dynamically ordered vortex lattice. We analyze the final structures and the behavior of the total energy of the system.


\end{abstract}

The phase diagram of vortices in high temperature superconductors in the presence of pinning potentials  has a rich variety of phases and transitions between them\cite{Nori,Laguna}. 
Several techniques have been implemented to artificially create pinning centers\cite{tecnicas}. In particular, the structure of the vortex system in the presence of a superficial pinning potential has been studied by means of Bitter decorations\cite{Fasano}.
Experiments show that there is no observable change in the critical current (the minimal current needed to depin the vortex lattice) after the decoration is performed. This indicates that the bulk pinning dominates the vortex dynamics. In a previous work we show that, for a set of parameters that are consistent with the experimentally observed structures, the depinning  force is dominated by the bulk pinning\cite{Laguna2}.   
Here we present additional results in which a sequence of numerical experiments have been performed to show that: 
a) When the external force exceeds the depinning force, the vortex lattice becomes dynamically ordered, even with a quasi-periodical Bitter pinning. This order is similar to the one predicted for a vortex system displaced by high forces over a random potential\cite{Koshelev}.
b) For a wide range of bulk pinning, polycrystalline structures are obtained after a gradual warming up from the dynamically ordered vortex lattice or by cooling down the system from high temperatures. 
c)These final structures are not reproducible due to the dense bulk pinning that randomly nucleates the solid phase. In spite of that, these low temperature structures have the same energy and melting temperature for the value of bulk pinning force used in this paper.
Based on these results, we conclude that there is a numerically efficient
way to treat the random bulk pinning present in the samples.
This allows us to cover a wide range of magnetic fields and to study the interplay between the weak surface and bulk pinning potentials.

As in previous works, we model the vortex system as a set of point-like particles interacting with pinning potentials and an effective vortex-vortex interaction\cite{Laguna2}. We perform numerical simulations with two-dimensional Langevin dynamics. 
The dynamical equations are 
$\eta {\mathbf v}_{i} = {\mathbf f}_{i} = {\mathbf f}_{i}^{vv} + {\mathbf f}_{i}^{vp} + {\mathbf f}_{i}^{T}+ {\mathbf f}_{i}^{b}$, where $ {\mathbf v}_{i} = d {\mathbf r}_{i}/dt$ 
is the vortex velocity and $\eta$ is the Bardeen-Stephen friction. The logarithmic vortex-vortex interaction $ {\mathbf f}_{i}^{vv}$ has a cut off, the Bitter pinning ${\mathbf f}_{i}^{vp}$ is modeled as a set of attractive Gaussian wells and the effect of temperature ${\mathbf f}_{i}^{T}$ is added as a stochastic term\cite{Laguna}.
To model the bulk pinning force ${\mathbf f}_{i}^{b}$ we compute the force acting on the {\it i}th vortex, which is allowed to move if the force is higher than a critical force $F_{c}$. We take $F_c$ time independent and uniform through all the sample.
We simulate a magnetic decoration experiment by gradually cooling down a fixed number of vortices $N_{v}$. 
To simulate the double decoration experiment, we cool down the same number of vortices in the presence of a pinning potential located at the vortex positions of the first decoration. We take $N_{v}=1024$ in all the results of this paper.
We simulate a dynamical ordering experiment adding to our system a homogeneous external force.

\begin{figure}[ht]
\vspace{-.6cm}
\includegraphics[width=.8\linewidth]{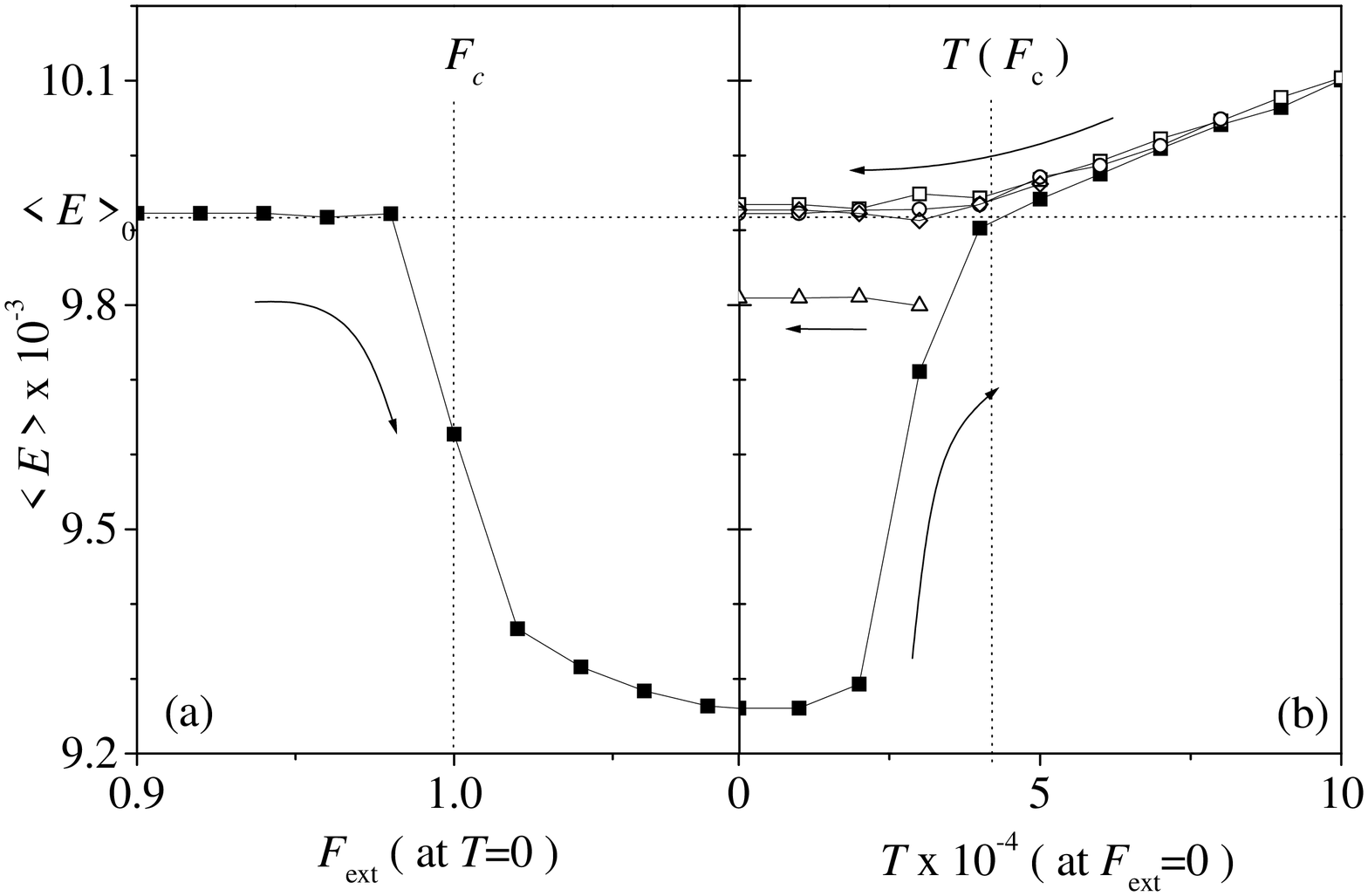} 
\vspace{-.7cm}
\caption{Average energy of the vortex system. (a) An external force is applied at $T=0$ until the lattice becomes ordered. $\left< E \right>_0$ is the energy before the application of $F_{ext}$. (b) The temperature is slowly varied at $F_{ext}=0$, starting from the dynamically ordered structure. Arrows indicate the direction of the variation in temperature.}  
\label{fig1} 
\end{figure}

\begin{figure}[ht]
\vspace{-.8cm}
\resizebox{\columnwidth}{!}{\includegraphics[80,520][550,720]{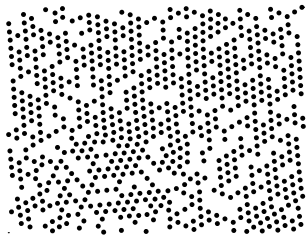}}
\vspace{-1.1cm} 
\caption{(a) Vortex system dynamically ordered. (b) Voronoi construction.}  
\label{fig2} 
\end{figure}

We perform a sequence of numerical experiments on a vortex system with a bulk pinning that leads to a polycrystalline low temperature structure. In our units, this is obtained with $F_{c} = 1$. 
In a first stage we cool down the vortex system on top of the first decoration, simulating a double decoration experiment.
In a second stage, we simulate a dynamical ordering experiment at $T=0$.
We calculate the average total energy $\left< E \right>$ of the vortex system in the stationary state. Note that the bulk pinning is so dense that its contribution is structure independent.
 In Fig. \ref{fig1} (a) we plot $\left< E \right>$ as function of the external force $F_{ext}$. As a consequence of the lattice ordering, the elastic contribution and the total energy $\left< E \right>$ decrease. In the picture we do not observe an energy increase caused by de depinning of the vortices, because the Bitter pinning contribution to the energy is one order of magnitude lower than vortex-vortex interaction one.
The decreasing of $\left< E \right>$ indicates a better ordered lattice. This coincides with the a reduction of the number of defects. The dynamically ordered lattice is shown in Fig. \ref{fig2} (a). To study the topological order of the vortex solid, we make Voronoi cell constructions, where the gray (black) polygons corresponds to vortices with five (seven) nearest neighbors, and the white ones corresponds to vortices with coordination number equal to 6 (see Fig. \ref{fig2} (b)). Before the application of the external force, the vortex system had several grains at random orientations, a defect fraction of $0.18$ and a high percentage of vortices at the Bitter pinning centers ($\sim 70 \%$).
The vortex lattice after ordered has a very low defect fraction ($< 0.02$). The final state may depend on the direction of the external force due to finite size effects: The dynamically ordered lattice tend to be oriented along the force direction. However the sample is commesurated with a given orientation of the triangular lattice. When $F_{ext}$ is along the $y$ direction, the final state has typically a single grain closed to the optimum orientation. For $F_{ext}$ along the $x$ direction, the boundary condition leads to structures with either two grains (like in Fig \ref{fig2}) or a single grain with crystalline axis forming a random angle with the direction of $F_{ext}$. 

Fig. \ref{fig1} (b) shows the third stage of the experiment. The starting point is the vortex lattice dynamically ordered. We turn off the external force and increase the temperature from zero. When the temperature is low, the system remains frozen by the action of the bulk pinning. At a higher temperature, a few vortices start to move and get trapped at the Bitter pinning centers, improving the overlap with the first decoration. The deformation of the originally perfect elastic lattice gives rise to a increase in the energy $\left< E \right>$. When the temperature reaches a characteristic value $T \sim T(F_{c})$, the vortex system depins. At this temperature the energy $\left< E \right>$ coincides with $\left< E \right>_0$, the energy of the system at the initial time.
If the system is cooled down from temperatures higher than $T(F_{c})$, the system energy converges to the original value $\left< E \right>_0$ (see Fig. \ref{fig1} (b)). 
Conversely, if the system is cooled down from temperatures lower than $T(F_{c})$, the energy is keeping almost fixed, as can be observed in Fig. \ref{fig1} (b) (open triangles).

In summary: 
{\it i}) 
When the system freezes, it has a characteristic energy $\left< E \right>_0$ corresponding to a wide set of metastable states. These states have a characteristic density of defects and overlap with the quasiperiodic Bitter pinning.
{\it ii}) 
An external force can order the vortex structure decreasing its energy due to a reduction of the vortex lattice defects. When the first and second decorations are dynamically ordered, the overlap between them is not systematically improved for the range of parameters used. 
{\it iii}) 
The dynamically ordered lattices have a high orientational order which depends on the direction of $F_{ext}$.

We thank F. de la Cruz, Y. Fasano and M. Menghini for helpful discussions. We acknowledge financial support from CONICET, CNEA, Fundaci\'on Antorchas and ANPCyT Grant No. 99-6343.

\end{document}